\documentclass[11pt,a4paper]{article}
\usepackage[hyperref]{acl2019}
\usepackage[pdftex]{graphicx}
\usepackage{times}
\usepackage{latexsym}
\usepackage{subcaption}
\usepackage{textcomp}

\usepackage{url}
\usepackage{color}

\usepackage{siunitx}
\aclfinalcopy % Uncomment this line for the final submission
\title{The merits of Universal Language Model Fine-tuning for Small Datasets -- a case with Dutch book reviews}

\author{Benjamin van der Burgh\\ {\small Leiden Institute of Advanced Computer Science} \\
   {\small Leiden University} \\
  {\small \texttt{b.van.der.burgh@liacs.leidenuniv.nl}}   \And Suzan Verberne\\
   {\small Leiden Institute of Advanced Computer Science} \\
   {\small Leiden University} \\
  {\small \texttt{s.verberne@liacs.leidenuniv.nl}} \\}

\date{}

\begin{document}
\maketitle
\begin{abstract}
% 1. what did we do
We evaluated the effectiveness of using language models, that were pre-trained in one domain, as the basis for a classification model in another domain: Dutch book reviews.
% 2. why did we do it
Pre-trained language models have opened up new possibilities for classification tasks with limited labelled data, because representation can be learned in an unsupervised fashion.
% 3. how did we do it
In our experiments we have studied the effects of training set size (100--1600 items) on the prediction accuracy of a ULMFiT classifier, based on a language models that we pre-trained on the Dutch Wikipedia.
We also compared ULMFiT to Support Vector Machines, which is traditionally considered suitable for small collections.
% 4. what did we find
We found that ULMFiT outperforms SVM for all training set sizes and that satisfactory results (\texttildelow 90\%) can be achieved using training sets that can be manually annotated within a few hours.
% 5. what do we think it means
We deliver both our new benchmark collection of Dutch book reviews for sentiment classification as well as the pre-trained Dutch language model to the community.
\end{abstract}

\section{Introduction}
Typically, results for supervised learning increase with larger training set sizes. However, many real-world text classification tasks rely on relatively small data, especially for applications in specific domains. Often, a large, \textit{unlabelled} text collection is available to use, but \textit{labelled} examples require human annotation. This is expensive and time-consuming. Since deep and complex neural architectures often require a large amount of labeled data, it has been difficult to significantly beat the traditional models -- such as Support Vector Machines -- with neural models~\cite{adhikari2019rethinking}. %Traditionally, the most-used text classification model was Linear Support Vector Machines (SVM), and still it is successful in many tasks where the amount of labelled data is limited or the computing facilities are limited. 

In 2018, a breakthrough was reached with the use of pre-trained neural language models and transfer learning~\cite{howard2018universal,peters2018deep,devlin2018bert,liu2019linguistic}. Transfer learning no longer requires models to be trained from scratch but allows researchers and developers to reuse features from models that were trained on different, much larger text collections (e.g. Wikipedia). For this pre-training, no explicit labels are needed; instead, the models are trained to perform straightforward language modelling tasks, i.e. predicting words in the text. 

Even though the models are trained on these seemingly trivial predictive tasks, transfer learning with these models is highly effective: the pre-trained language models can be fine-tuned to perform classification tasks with a relatively small amount of labelled task-specific data. Thus, pre-trained language models can alleviate the small labelled data size for domain-specific data sets. 

In their 2018 paper, Howard and Ruder show the success of transfer learning with Universal Language Model Fine-tuning (ULMFiT) for six text classification tasks. They also demonstrate that the model has a relatively small loss in accuracy when reducing the number of training examples to as few as 100~\cite{howard2018universal}.

In this paper we further address the use of ULMFiT for small training set sizes. We consider the case of data from a new domain, where we have a large amount of unlabelled data, but limited labelled data. Given the vast number of network parameters and the limited number of training instances (100 to 1600), we expect to quickly overfit on the training data if all parameters are optimized using the small labelled data, often referred to as \textit{catastrophic forgetting}.
Alternatively, we `freeze' the parameters of the language model, which means that we fix all network parameters except for the parameters of the final layer.
In doing so, we limit the ability of the model to adapt to the target domain, but in return avoid the problem of catastrophic forgetting, since the language model parameters are untouched.

In this paper, we evaluate ULMFiT with a pre-trained language model and fixed hyperparameters for the representation layers.
We only tune the drop-out multiplier and learning rate for the linear layers.%, and investigate the differences between the optimal parameter values on different random samples of small training collections.

For our experiments on Dutch texts, we created a new data collection consisting of Dutch-language book reviews. We fine-tune a general pre-trained Wikipedia model on the reviews collection. We then take various-sized labelled portions of the book review data to (a) investigate the effect of training set size, and (b) compare the accuracy of ULMFiT to the accuracy of Support Vector Machines (SVM). %We address the following research questions: \begin{enumerate} \item To what extent are the default parameter settings generalizable to smaller training set sizes? \item To what extent is the quality of ULMfit robust with smaller training set sizes, and how does this quality compare to Support Vector Classification? \end{enumerate}

The contributions of this paper compared to previous work are: (1) We deliver a new benchmark dataset for sentiment classification in Dutch; (2) We deliver pre-trained ULMFiT models for Dutch language; (3) We show the merit of pre-trained language models for small labeled datasets, compared to traditional classification models. %\blue{Moet in deze contributions nog iets over hyperparameters?} \red{B: Dat zou makkelijker zijn als we een vergelijking maken met bijv. default parameters.}

\section{Data}
\paragraph{Data set} We released the 110k Dutch Book Reviews Dataset (110kDBRD).\footnote{\url{https://benjaminvdb.github.io/110kDBRD/}, also including the scripts used to scrape the data from the review website.} This dataset contains book reviews along with associated binary sentiment polarity labels. It is inspired by the Large Movie Review Dataset~\cite{maas2011} and intended as a benchmark for sentiment classification in Dutch. We scraped 110 thousand book reviews from the website Hebban.\footnote{\url{https://www.hebban.nl}} These reviews each consist of a text and a score from 1 to 5, which we converted to categorical labels (1 and 2: negative; 3: neutral; 4 and 5: positive).

\paragraph{Data split} For our experiments, we split the 110k documents as follows: 20k documents are used in the classifier training and evaluation (positive and negative classes balanced). Of those 20k, we reserve 5k documents as a held-out test set that cannot be consulted during training nor language model pre-training. The remaining 15k documents are used for training the classifier.

\paragraph{Data sampling for training} For the experiments on dataset size we use the following training set sizes $m=\{ 100, 200, 400, 800, 1600\}$. Each experiment is trained 10 times to investigate model stability. These 10 subsamples are chosen randomly out of the complete 15k training set (not balanced). Note that the same test set is used for all experiments to make the results directly comparable.

\section{Language model training}

\subsection{General-domain language model pre-training}

In order to learn text representations we use the AWD-LSTM language modelling architecture originally used by ULMFiT and implemented in the Fast.ai Python library.
This library also constructs a supervised dataset by randomly masking out words in a text.
We use a unidirectional language model, i.e., the target word is predicted using words that precede it.
Similarly to \citet{howard2018universal}, we have chosen Wikipedia for language modelling, because it provides a large, freely available corpus of high quality.
Moreover, we found that the pre-processing scripts for the English version of Wikipedia could be re-used for the Dutch.

We used a recent dump of Wikipedia and converted it to raw text, which was then split on white spaces into tokens. After that, we replaced all numbers with the same placeholder token, such that the specific value is ignored, but the fact that a number occurred can be used in the model.
The 60k most frequent tokens were included in the vocabulary $V$ and the remaining out-of-vocabulary words were replaced with a special `unknown' token.
An embedding layer of size 400 was used to learn a dense token representation, followed by 3 LSTM layers with 1150 hidden units each to form the encoder.

This is followed by a classification module that maps each representation to a score $0 \leq s_t \leq 1$, for each token $t \in V$ where $\sum{s_t} = 1$ and can as such be interpreted as a probability distribution over the vocabulary.
We used the reference implementation of ULMFiT in the Fast.ai Python library.
%We assume that the dataset mainly contain high-quality articles and that low-quality or duplicate articles will not greatly impact the language model.
The entire Wikipedia dataset was split into 92M tokens for training and 185k for both testing and validating the language model.
A slanted triangular learning rate \cite{howard2018universal} with a learning rate of $5*10^{-3}$ was used for 20 epochs.

\subsection{Target task language model fine-tuning}

After training the language model on Wikipedia, we continued training on data from our target domain, i.e., the 110k Dutch Book Review Dataset. The preprocessing was done similarly to the preprocessing on Wikipedia, but the vocabulary of the previous step was reused. We used all data except for a 5k holdout set (105k reviews) to fine-tune network parameters using the same slanted triangular learning rates.
However, this time we first trained the parameters of the classification module to convert the pre-trained features into predictions for the new target dataset.
After that all network parameters were trained for 10 epochs.

\subsection{Target task classifier fine-tuning}

The goal is to predict the sentiment polarity (positive or negative) given a review text.
Therefore, the training dataset is constructed such that the dependent variable represents a sentiment polarity instead of a token from the vocabulary.
The encoder of the language model is kept, such that a dense representation can be constructed given an input text, and the classification module is replaced to adjust for the new target classes.

\section{Experiments}\label{experiments}

\subsection{Preprocessing}
%We compare two tokenization settings: (a) regular word-based tokenization with Spacy\footnote{\url{http://spacy.io}}; (b) SentencePiece, a language-independent subword tokenizer designed for Neural-based text processing~\cite{kudo2018sentencepiece} --> niet gedaan

We applied the default text processing implemented in the Fast.ai Python library by splitting on whitespace and padding texts within a batch to the same length.
The amount of required padding characters was reduced by grouping texts of similar length together, while adding some randomness during training to avoid showing the network with the same batches in each epoch.

\subsection{Hyperparameters}
%First, we investigate how large the difference is between the optimal hyperparameters for an extremely small training set size (2 categories, 200 items per category). Second, we investigate what the difference is in classifier quality when we use either the default parameters (determined on a large training set size) or the optimal parameters for a small data set.
We optimized hyperparameters for each training set size and for each fold using HpBandster.\footnote{\url{https://github.com/automl/HpBandSter}} A one-cycle policy, as outlined in \cite{smith2018}, was used, which requires a lower and upper bound for the momentum, describing its adaptive curve during a single epoch. This resulted in five optimized hyperparameters: learning rate, momentum lower and upper, dropout and batch size. In the objective function, we optimized for binary cross-entropy loss.

%15 minutes * 6 * 10 = 15 uur

%90\% train, 10\% test om optimaliseren op elke random set voor elke training set grootte. Surrogaatmodel = Bayesian model. Redelijk noisy estimates, random sampelen. Vaak doen. Budget sizes. Eerst klein budget, kijken of het een interessante combinatie is. Als er genoeg evidence is in deze omgeving dan meer budget. Aantal epochs eerst 4, dan uitbouwen. Dat draait nu voor de grootte 1000 (1x).

%def get_configspace():
%        cs = CS.ConfigurationSpace()
%        lr = CSH.UniformFloatHyperparameter('lr', lower=1e-5, upper=1e-1, default_value='1e-2', log=True)
%        moms1 = CSH.UniformFloatHyperparameter('moms1', lower=0.6, upper=0.99, default_value=0.8, log=False)
%        moms2 = CSH.UniformFloatHyperparameter('moms2', lower=0.5, upper=0.99, default_value=0.7, log=False)
%#         bs = CSH.UniformIntegerHyperparameter('bs', lower=1, upper=32, default_value=16)
%        drop_mult = CSH.UniformFloatHyperparameter('drop_mult', lower=0.1, upper=0.9, default_value=0.5)
%        cs.add_hyperparameters([lr, moms1, moms2, bs, drop_mult])
%        return cs

\subsection{Baselines}
We compared our classification models to Linear Support Vector Machines (SVM) because it is a commonly used and well performing classifier for small text collections. We used the implementation of LinearSVC in scikit-learn.\footnote{\url{https://scikit-learn.org/stable/modules/generated/sklearn.svm.LinearSVC.html}} LinearSVC has one hyperparameter, C, which we optimized using HpBandster on the range of values from $10^{-4}$ to $10^4$ with squared hinge loss as optimization function (default for LinearSVC in scikit-learn). For feature extraction we used the CountVectorizer and TF-IDF transformer in scikit-learn. TF-IDF weights were trained on the same 105k documents on which the ULMFiT model was fine-tuned.

For comparison we also trained two models, one SVM and one ULMFiT model, with manually tuned hyperparameters on all available book reviews in the training set (15k). These models achieved 93.84\% (ULMFiT) and 89.16\% (SVM).

\section{Results}
\begin{figure*}[th]
    \centering
\begin{minipage}[t]{0.50 \textwidth}
  \centering
  \includegraphics[scale=0.58]{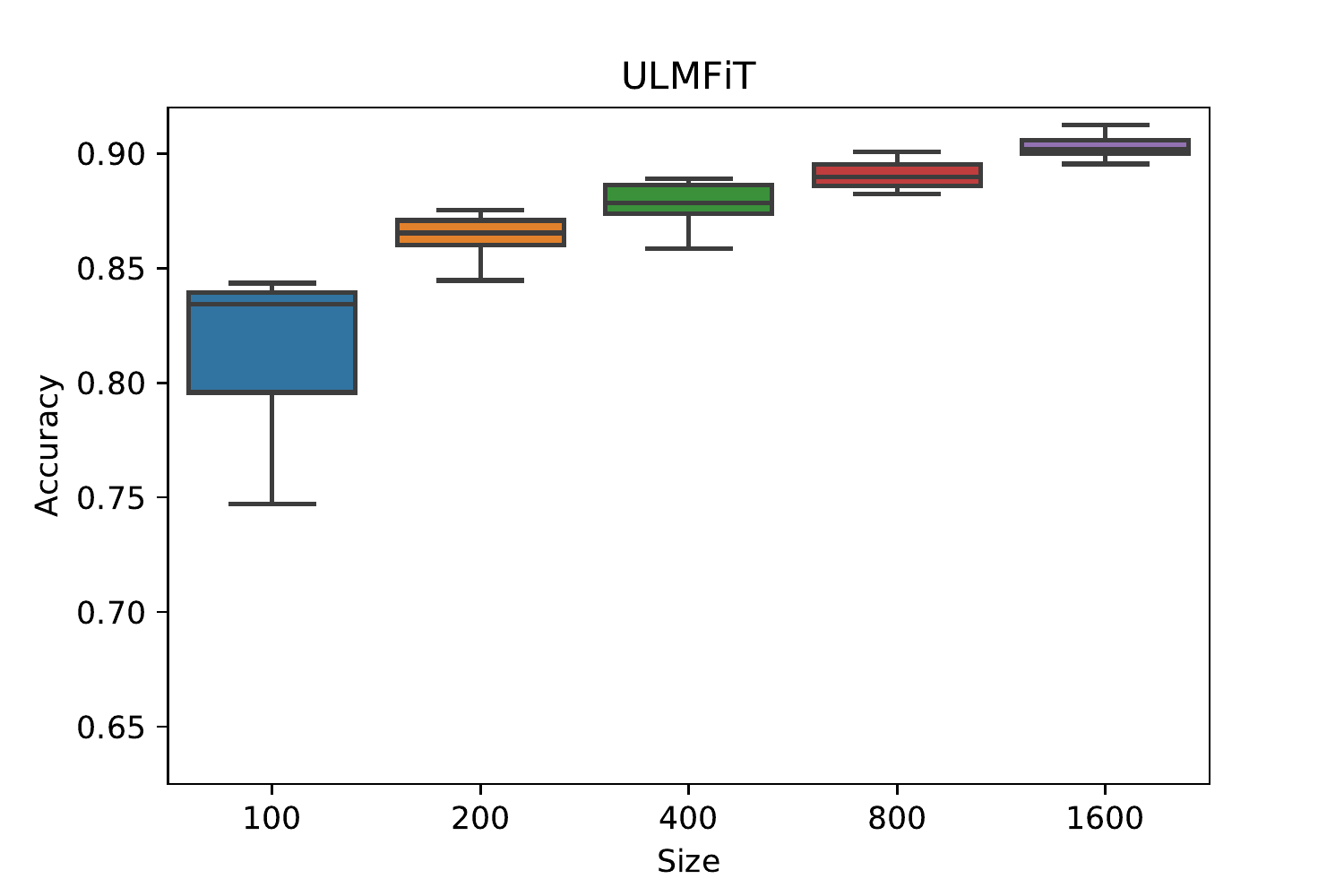}
\end{minipage}%
\begin{minipage}[t]{0.50 \textwidth}
  \centering
  \includegraphics[scale=0.58]{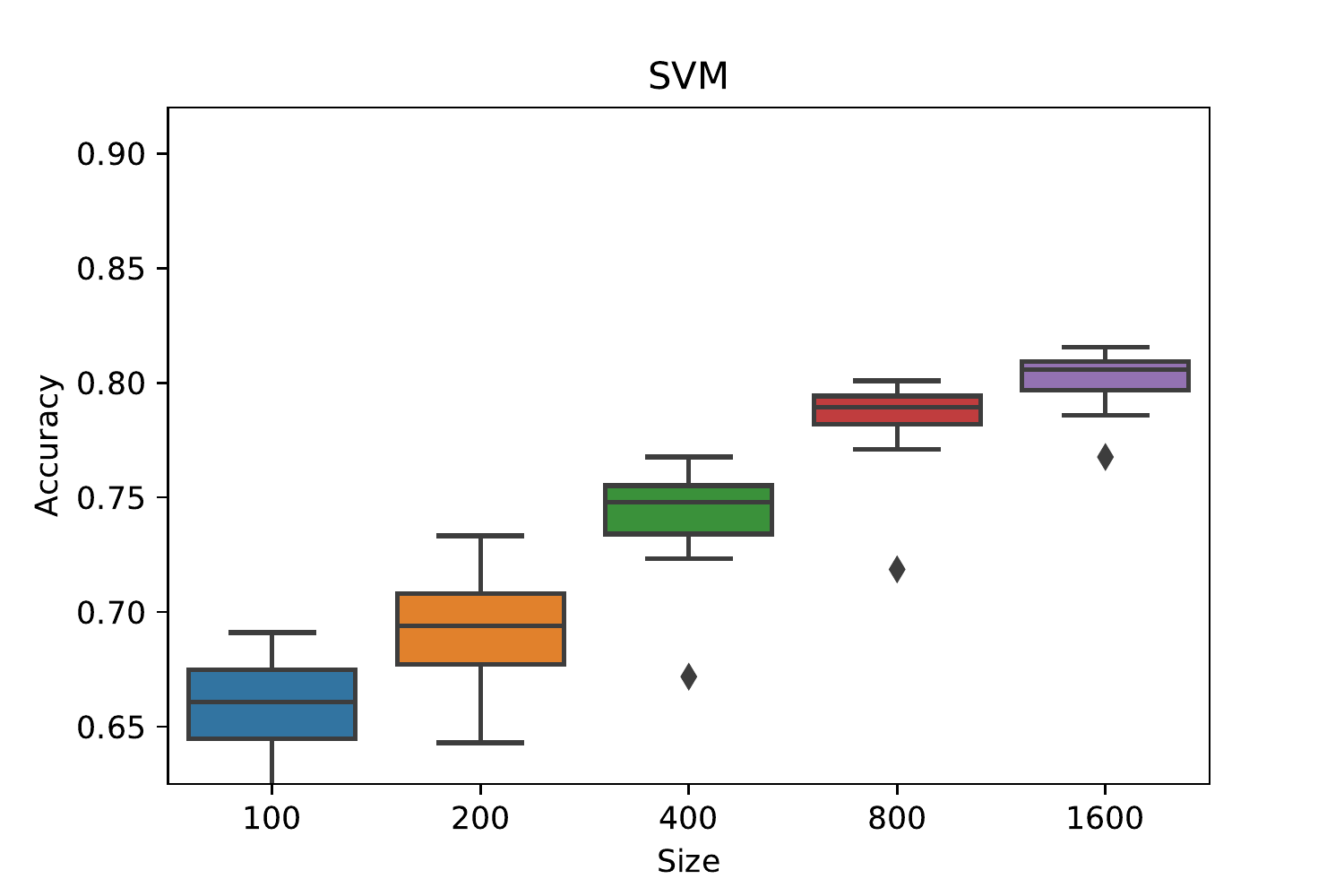}
\end{minipage}%
    \caption{Results for ULMFiT (a) and SVM (b) in terms of accuracy on the test set with varying training set sizes. The boxes represent the deviation among the random subsample per training set size.}
    \label{fig:boxplot}
\end{figure*}

%\subsection{Classification results}
% \begin{table}[t]
%     \centering
%     \begin{tabular}{lr}
%     \hline
% ULMFiT with pre-trained LM & 93.84\%\\
% ULMFiT no pre-train & 92.55\%\\
% SVM & 89.16\%\\
% Flair with fastText & 88.48\%\\
% fastText & 80.90\%\\
% \hline
%     \end{tabular}
%     \caption{Classification results (accuracy) on the complete 110kDBRD dataset. \blue{Zijn dit oude resultaten? Gaan we die eruit halen? Of alleen de 1e drie rijen erin laten?}}
%     \label{tab:mainresults}
% \end{table}

\subsection{Effect of training set size}

The results of the experiments described in Section~\ref{experiments} can be found in Figure~\ref{fig:boxplot} (left). A few observations can be made from this plot. Firstly, for the ULMFiT model, the accuracy on the test set improves with each increase in the training dataset size, as can be expected.

Secondly, both models behave rather unstable for smaller training datasets, as can be seen by the large deviation from the mean and the outliers: different random subsamples give deviant results for the smaller training set sizes.
Since the pre-trained model is based on data from a different domain, it can be expected that more than 100 instances are needed to accommodate for the new domain.

\subsection{Comparison to SVM}

Figure~\ref{fig:boxplot} compares the prediction accuracies for ULMFiT and SVM. %shows that the prediction accuracy of the ULMFiT model shows a steady increase with increasing training set sizes, while the SVM model seems to converge.
We had expected the SVM model to perform better for smaller training set sizes, but it is outperformed by ULMFiT for each size. Also, the ULMFiT models show smaller deviations between random subsamples than the SVM models.

We also found that the prediction accuracy of the SVM model using all 15,000 training items (89.16\%) is surpassed by the ULMFiT model when using only 1600 training instances: all 10 random subsamples for ULMFiT reach an accuracy of at least 89.54\% (the left purple box in Figure~\ref{fig:boxplot}).
This could mean that the pre-trained model captures many of the required characteristics of Dutch such that they can be largely used without modifications.
%It is more interesting is that the number of instances required to approach the accuracy of the model that used all available data (\texttildelow 94\%) is not as high as might be expected.

\section{Conclusions}

%A common argument against the use of deep neural networks is often the required amount of labeled data. 
Pre-trained language models have opened up possibilities for classification tasks with limited labelled data.
In our experiments we have studied the effects of training set size on the prediction accuracy of a ULMFiT classifier based on pre-trained language models for Dutch. In order to make a fair comparison, we have used state-of-the-art optimization methods to optimize the hyperparameters of each model.

Our results confirm what had been stated in \cite{howard2018universal}, but had not been verified for Dutch or in as much detail. For this particular dataset and depending on the requirements of the model, satisfactory results might be achieved using training sets that can be manually annotated within a few hours.

Moreover, a large part of modeling effort lies in the training of a language model on an -- in this case -- generic corpus, which can be reused for other domains. 
While the prediction accuracy could be improved by optimizing all network parameters on a large dataset, we have shown that training only the weights of the final layer outperforms our SVM models by a large margin.

ULMFiT uses a relatively simple architecture that can be trained on moderately powerful GPUs.
This fact, combined with the general availability of unlabeled data and the ability to share language models, suggests that these methods could be applied in domains where manual labelling has traditionally been too expensive.
Further research should be conducted to compare how differences between the source and target datasets affect the prediction accuracy and whether more powerful network architectures can also be used.

%\section*{Acknowledgments}

%The acknowledgments should go immediately before the references.  Do not number the acknowledgments section. Do not include this section when submitting your paper for review. \\

%\bibliography{ulmfit}
\bibliographystyle{acl_natbib}

\end{document}